\documentclass[prB,twocolumn,letterpaper,showpacs]{revtex4}              
\usepackage{graphicx}
\usepackage{pslatex}
\usepackage{amsmath}
\usepackage{amssymb}
\usepackage{graphicx}
\usepackage{pstricks}
\usepackage{wasysym}

\begin{document}
\bibliographystyle{apsrev}

% Use the \preprint command to place your local institutional report
% number on the title page in preprint mode.
% Multiple \preprint commands are allowed.
%\preprint{}

%Title of paper

\title{Hysteretic phenomena in many-body physics}

% Optional argument for running titles on pages
%\title[]{}

\author{Ivan Danchev}
\affiliation{Department of Physics and Astronomy, University of South Carolina, Columbia, SC 29208}

%\homepage[]{Your web page}
%\thanks{}
%\altaffiliation{}
\date{\today}

\begin{abstract}

We show that the cyclic evolution of an order parameter in a many-body system that has spontaneous symmetry breaking in the ground state brings about a dissipative geometric phase. This phase originates from the same mechanism that leads to dissipation in thermo-field dynamics studied earlier by Umezawa and his collaborators and appears to be a manifestation of the dissipative geometric phase introduced recently by Jain and Pati. 

\end{abstract}
% insert suggested PACS numbers in braces on next line
\pacs{75.40.Cx,73.22.Gk,75.60.Ej,03.65.Vf}
%\maketitle must follow title, authors, abstract and \pacs
\maketitle
%%%%%%%%%%%%%%%%%%%%%%%%%%%%%%%%%%%%%%%%%%%%%%%%%%%%%%%%%%%%%%%%%%%

\section{Introduction}

Recently Jain and Pati \cite{JainPati} have shown that dissipation in many-body systems leads to certain geometric phases. It is tempting to examine more closely the conditions associated with the appearance of those phases. It is well known that the low-energy response of many-body systems with spontaneously broken symmetries is dominated by gapless Nambu-Goldstone (NG) bosons. The fact that in Jain and Pati's approach the geometric phase is related to the static susceptibility and hence to the low-energy response of the system is suggestive of a possible role played by spontaneous symmetry breaking. 

Many-body systems with a broken symmetry of the ground state have certain order parameters and it is natural to expect that the response of a time-dependent external perturbation that couples to an order parameter will cause its change in time. Then, following Berry's original argument \cite{Berry}, one might expect that a cyclic evolution in the (order) parameter space can result in a geometric phase. Since order parameters in many-body systems are generated by symmetry breaking, no additional parametric dependence of the Hamiltonian is needed. 

It is also well known that the Fock space of a many-body system is not unique but splits into folia of inequivalent representations, each labeled by order parameters arising from spontaneous symmetry breaking \cite{UMWA1}. This implies that any temporal evolution of a many-body system involving change of an order parameter takes place within a certain class of inequivalent Fock space representations. Umezawa and his collaborators working in the formalism of thermo-field dynamics \cite{UMWA1}, \cite{UMWA2} have already shown that temporal evolution through thermally inequivalent vacuua gives rise to dissipation, and Vitiello \cite{Vitiello} has specifically pointed out that a geometric phase must be associated with it.

All this taken together provides a motivation to examine the conditions that lead to a dissipative geometric phase in a many-body system with spontaneous symmetry breaking. We investigated here these conditions using, as an example, the Stoner model of itinerant electron ferromagnetism \cite{Stoner}. This model has been extensively studied in the past \cite{HerringKittel},\cite{Moriya}, and various aspects of the mechanism of spontaneous symmetry breaking in it have been elucidated \cite{Mattuck},\cite{MUST}. Recently it was applied in the study of isospin excitations in nuclei \cite{Danchev}.
Since, to the best of our knowledge, there exist no previous attempts to relate a spontaneously broken symmetry to dissipation and the geometric phase, we consider it worthwhile to study such a relation for a particular case of the Stoner model. 
The results obtained here appear to be generic to systems with broken symmetry.

The rest of this paper is organized as follows.

In Section II we briefly review the main features of the Stoner model at zero temperature following a variational approach which gives the equation for the order parameter (magnetization) from  minimization of the ground state energy of the system. Next, in Section III we show how the emergence of order in the system is related to the appearance of a gapless collective mode, and hence to the low-energy response in the system. Both sections are in the nature of an abridged review, and the reader is referred to the original articles of Mattuck \cite{Mattuck} and Matsumoto et al.\cite{UMWA1} for details. In Section IV we study the closed contour integral of magnetization (the order parameter) as a function of an the applied external field (time dependent cyclic perturbation) and show that it is non-zero if the system is in the symmetry broken phase, and zero otherwise. That essentially gives rise to the dissipative geometric phase of Jain and Pati in this system and leads to the well known hysteresis curve in magnetic systems \cite{Bertotti}. Finally, we extend our results to finite temperature.

%%%%%%%%%%%%%%%%%%%%%%%%%%%%%%%%%%%%%%%%%%%%%%%%%%%%%%%%%%%%%%%%%%%%%%%%%%%%%%%%%%%%%%%%%%%%%%%
%
%               SECTION 2
%
%%%%%%%%%%%%%%%%%%%%%%%%%%%%%%%%%%%%%%%%%%%%%%%%%%%%%%%%%%%%%%%%%%%%%%%%%%%%%%%%%%%%%%%%%%%%%%%
\section{Stoner model of ferromagnetism}
\label{sec:Stoner model}

The Hamiltonian of a non-relativistic many-body theory of interacting fermions is generically given by
\begin{eqnarray}
H&=&\int d^3x\psi^{\dagger}(x)(\frac{-\hbar^2\nabla^{2}}{2m})\psi(x)\nonumber\\
& +&\frac{1}{2}\int d^{3}xd^{3}y\psi^{\dagger}(x)\psi^{\dagger}(y)V(x-y)\psi(y)\psi(x).
\end{eqnarray}
Here $\psi$'s are Pauli spinors and the two-body interaction potential is SU(2)-spin invariant
\begin{equation}
V(x-y)=V_{0}(\mid x-y\mid)+V_{1}(\mid x-y\mid)\sigma_{1}\cdot\sigma_{2}.
\end{equation}
In the Stoner model this SU(2)-spin invariant electron-electron interaction is treated schematically as a contact-type interaction with a fixed strength. 
\begin{equation}
V(x-y)=\lambda\delta^{(3)}(x-y).
\end{equation}
The interaction Hamiltonian reduces then to 
\begin{equation}
V=\lambda\int d^{3}
 x\psi_{\uparrow}^{\dagger}(x)\psi_{\downarrow}^{\dagger}(x)
  \psi_{\downarrow}(x)\psi_{\uparrow}(x).
\end{equation}
The densities of spin-up and spin-down electrons are related to their respective Fermi momenta as 
\begin{equation}
n_{\uparrow}=\frac{(k_{F}^{\uparrow})^{3}}{6\pi^{2}} \hspace{1in}
n_{\downarrow}=\frac{(k_{F}^{\downarrow})^{3}}{6\pi^{2}}.
\end{equation}
While the total electron density $n$ remains unchanged, 
\begin{equation}
\frac{k_{F}^{3}}{3\pi^{2}}=n=n_{\uparrow}+n_{\downarrow}=\frac{(k_{F}^{\uparrow})^{3}}{6\pi^{2}}+\frac{(k_{F}^{\downarrow})^{3}}{6\pi^{2}},
\end{equation}
the interaction may change the individual densities of spin-up and spin-down electrons, giving non-zero magnetization $M$
\begin{equation}
M=n_{\uparrow}-n_{\downarrow}.
\end{equation}
It is also useful to define the Fermi energy and momentum of unpolarized electron gas by
\begin{equation}
\label{Fermiun}
\epsilon_{F}=\frac{\hbar^{2}k_{F}^{2}}{2m}=\frac{\hbar^{2}(3\pi^{2}n)^{2/3}}{2m}.
\end{equation}
We now evaluate the ground state energy of the system. The kinetic term is given by
\begin{equation}
\langle T\rangle=\frac{\hbar^2}{20m\pi^{2}}(6\pi^{2}n_{\uparrow})^{5/3}+\frac{\hbar^2}{20m\pi^{2}}(6\pi^{2}n_{\downarrow})^{5/3}.
\end{equation}
Defining a dimensionless parameter $\Delta$, describing spin polarization (magnetization) by
\begin{equation}
\label{Delta}
\Delta=\frac{n_{\uparrow}-n_{\downarrow}}{n_{\uparrow}+n_{\downarrow}},
\end{equation}
one can express the $n_{\uparrow}$, $n_{\downarrow}$ and $M$ as functions of $\Delta$ 
\begin{eqnarray}
\label{definitions}
n_{\uparrow}=\frac{n}{2}(1+\Delta), \quad
n_{\downarrow}=\frac{n}{2}(1-\Delta), \quad
M=n\Delta.
\end{eqnarray}
Thus one arrives at 
\begin{equation}
\langle T\rangle=\frac{3}{10}n\epsilon_{F}
 [(1+\Delta)^{5/3}+(1-\Delta)^{5/3}].
\end{equation}
The potential energy contribution to the ground state energy in Hartree-Fock appoximation is written as 
\begin{eqnarray}
\langle V\rangle_{HF}= \lambda\langle\psi_{\uparrow}^{\dagger}(x)\psi_{\uparrow}(x)\rangle
                              \langle\psi_{\downarrow}^{\dagger}(x)\psi_{\downarrow}(x)\rangle \nonumber \\ 
=\lambda n_{\uparrow}n_{\downarrow}=\frac{\lambda n^2}{4}(1-\Delta ^2).
\end{eqnarray}
The ground state energy per electron in the Hartree-Fock approximation (in units of $\epsilon_{F}$) is
\begin{equation}
E(\Delta)=\frac{3}{10}[(1+\Delta)^{5/3}+(1-\Delta)^{5/3}]+\frac{\lambda n}{4\epsilon_F}(1-\Delta ^2).
\end{equation}
The value of $\Delta$ can be determined by minimizing the ground state energy.
\begin{equation}
0=\frac{\partial 
E(\Delta)}{\partial\Delta}=\frac{5}{10}[(1+\Delta)^{2/3}-(1-\Delta)^{2/3}]-\frac{\lambda n}{2\epsilon_{F}}\Delta.
\end{equation}
We arrive at the Stoner equation, which determines whether or not the system undergoes spontaneous magnetization \cite{Mattuck}
\begin{equation}
\label{Stoner}
\frac{\lambda n}{2\epsilon_{F}}=\frac{1}{2\Delta}[(1+\Delta)^{2/3}-(1-\Delta)^{2/3}].
\end{equation}
In FIG. 1, we display the plot of magnetization as a function of 
$x=\lambda n/(2\epsilon_{F})$.
%
%%%%%%%%%%%%%%%%%%%%%%%%%%%%%%%%%%%%%%%%%%%%%%%%%%%%%%%%%%%%
\begin{figure}[hbt]
\includegraphics[width=2.5in]{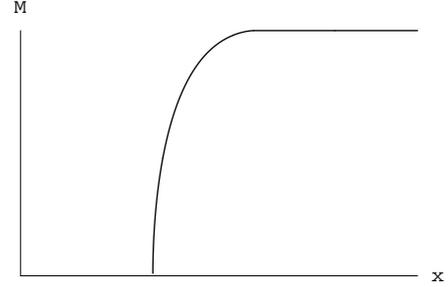}
\caption{Spontaneous magnetization as a function of the density and coupling constant\cite{Mattuck}}
\label{magnetization}
\end{figure}
%%%%%%%%%%%%%%%%%%%%%%%%%%%%%%%%%%%%%%%%%%%%%%%%%%%%%%%%%%%%
%
For $x\le 2/3 $, the system does not develop any spin polarization, i.e., it remains nonmagnetic. For $ 2/3< x < 2^{-1/3}$ the system becomes partially magnetized, and for $x>2^{-1/3}$ it is fully magnetized.
%\vspace{3in}
%%%%%%%%%%%%%%%%%%%%%%%%%%%%%%%%%%%%%%%%%%%%%%%%%%%%%%%%%%%%
\begin{figure}[hbt]
\includegraphics[width=2.5in]{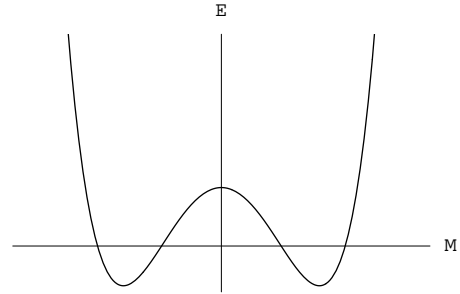}
\caption{Ground state energy of the Stoner model in Hartree-Fock -approximation as a function of magnetization\cite{Mattuck}}
\label{gs_magnetization}
\end{figure}
%%%%%%%%%%%%%%%%%%%%%%%%%%%%%%%%%%%%%%%%%%%%%%%%%%%%%%%%%%%%

To summarize, in the Stoner model when a certain combination of the density and the strength of the interaction reaches a critical value, the system of electrons displays a characteristic ground state instability with respect to spontaneous magnetization.(In this and following figures $M$ is used to label the axes, while the actual scale is given in terms of the dimensionless $\Delta$.)
%
%%%%%%%%%%%%%%%%%%%%%%%%%%%%%%%%%%%%%%%%%%%%%%%%%%%%%%%%%%%%%%%%%%%%%%%%%%%%%%%%%%%%%%%%%%%%%%%%%%%%%%%
%
%          Section 3               
%
%%%%%%%%%%%%%%%%%%%%%%%%%%%%%%%%%%%%%%%%%%%%%%%%%%%%%%%%%%%%%%%%%%%%%%%%%%%%%%%%%%%%%%%%%%%%%%%%%%%%%%%

\section { Magnetization and low-energy spin response} 
\label{sec:order parameter - NG boson}
As is well known, the appearance of an order parameter in a quantum many-body system implies the existence of a bosonic degree of freedom in the quasiparticle Fock space of this system. This general statement is known as the Goldstone theorem \cite{Goldstone}. 

Important as it is, Goldstone's theorem turns out to be just one identity out of a chain of the Ward-Takahashi\cite{Ward},\cite{Taki} (WT) identities between the exact Green's functions of the system. These identities show how the original symmetry of the dynamics of the Heisenberg fields is manifested in the properties of the quasiparticles \cite{UMWA1}. Here we recapitulate the derivation of these identities according to the paper of Matsumoto et al.\cite{MUST}.

In our non-relativistic field theory of interacting fermions with an $SU(2)$-spin invariant Lagrangian density, we add an external magnetic field, $h$, that couples to magnetization breaking explicitly the $SU(2)$ symmetry down to $U(1)$. 

\begin{equation}
L_{h}=L(x)+h\mu_{B}\psi^{\dagger}(x)\sigma_{3}\psi(x).
\end{equation}
Next, we perform a global phase transformation on the fermionic fields 
\begin{equation}
\psi(x)\rightarrow \exp[i{\mathbf \sigma}\cdot{\mathbf \alpha}]\psi(x),
\end{equation}
which we regard as a change of variables in the generating functional of the Green's function of our model
\begin{equation}
W[\eta^\dagger,\eta]_{h}=\frac{1}{N}\int D \psi D \psi^{\dagger} \exp\{ i\int d^4 x [L_{h}(x)
+\eta^{\dagger}\psi+\psi^{\dagger}\eta]\},
\end{equation}
where  $N\equiv W[0,0]$.
One then notes that, because the functional integral does not depend on the integration variables, the $\alpha$-dependence of $W_h$ is fictitious, and hence one should impose the following requirement 
\begin{equation}
\frac{\partial W_{h}}{\partial{\mathbf \alpha}}=0.
\end{equation}
From this condition follow the basic Ward-Takahashi identities   
\begin{widetext}
\begin{eqnarray}
\int d^4 x\langle\eta^{\dagger}(x)\sigma_{3}\psi(x)-\psi^{\dagger}(x)\sigma_{3}\eta(x)\rangle_{\eta\dagger,\eta,h}&=&0 \\
\int d^4 x\langle\eta^{\dagger}(x)\sigma_{+}\psi(x)-\psi^{\dagger}(x)\sigma_{+}\eta(x)\rangle_{\eta\dagger,\eta,h}&=&
-2\mu_{B}h\int d^4 x \langle\sigma_{+}(x)\rangle_{\eta\dagger,\eta,h}\\
\int d^4 x\langle\eta^{\dagger}(x)\sigma_{-}\psi(x)-\psi^{\dagger}(x)\sigma_{-}\eta(x)\rangle_{\eta^{\dagger},\eta,h}&=&
+2\mu_{B}h\int d^4 x\langle\sigma_{-}(x)\rangle_{\eta^{\dagger},\eta,h},
\end{eqnarray}
\end{widetext}
where $\sigma_{\pm}(x)\equiv\psi^{\dagger}(x)\sigma_{\pm}\psi(x)$ is the composite boson field and $\sigma_{\pm}=\sigma_1+i\sigma_2$.
By taking appropriate functional derivatives of the basic Ward-Takahashi identities with respect to the sources, $\eta, \eta^{\dagger}$ and then setting  these sources to zero, one obtains a chain of identities between the exact Green's functions in the theory. The lowest order identities in this chain are 
\begin{widetext}
\begin{eqnarray}
M_{h}(x)\equiv\langle\psi^{\dagger}(x)\sigma_{3}\psi(x)\rangle_{h}&=&-2\mu_{B}h\int d^4 y\langle\sigma_{+}(y)\sigma_{-}(x)\rangle_{h} \\
\langle\psi_{\uparrow}(z)\psi_{\uparrow}^{\dagger}(y)\rangle_{h}-\langle\psi_{\downarrow}(z)\psi_{\downarrow}^{\dagger}(y)\rangle_{h}&=&-2\mu_{B}h\int d^4 x\langle\psi_{\downarrow}(z)\psi_{\uparrow}^{\dagger}(y)\sigma_{-}(x)\rangle_{h}.
\end{eqnarray}
\end{widetext}
The first identity shows that the difference in the densities between the spin-up and the spin-down electrons, i.e. the order parameter induced by the external magnetic field $h$, is proportional to the space integral of the propagator of the composite boson field. This boson field is the magnon (spin-wave). It describes the transverse oscillations of the order parameter.
The second Ward-Takahashi identity relates the exact propagators of spin-up and down electrons to their effective coupling to the spin-wave given by the vertex function on the r.h.s of the identity. 
From our discussion in Section II we already know that for certain ranges of the values of the coupling constant and the density, non-zero magnetization may appear even in the absence of an external field. This means that the l.h.s of the first WT-identity may be non-zero in the limit $h\rightarrow0$, which in turn forces the r.h.s to be non-zero in that limit. In the momentum representation this identity is expressed as
\begin{eqnarray}
M=-\lim_{h\to 0}2\mu_{B}h \left(\frac{\chi(q^2)}{q_{0}-\omega({\mathbf q})+i\varepsilon}\right)_{q_{0}=0,{\mathbf q}=0},
\end{eqnarray}
where $\chi(q^2)$ is the boson (magnon)  spectral function, and $\omega(\mathbf{q})$ represents its dispersion relation $\omega=\omega({\bf q})$. This identity  implies that in the long wavelength limit, we should have $\lim_{{\mathbf{q}}\to 0}\omega({\bf q})=0$ and $\lim_{{\mathbf{q}}\to 0}\chi(0)=M$. In other words a non-zero magnetization in the absence of external field requires the existence of a gapless, NG boson field. This in essence is the Goldstone theorem in the context of non-relativistic quantum field theory. This NG boson is physically identified with spin waves or magnons in this system. Its propagator has a pole at zero momentum which gives the dominant, low-energy spin response of the system.
%
%%%%%%%%%%%%%%%%%%%%%%%%%%%%%%%%%%%%%%%%%%%%%%%%%%%%%%%%%%%%%%%%%%%%%%%%%%%%%%%%%%%%%%%%%%%%%%%%%%%%%%%%%%%%%%%
%
%         Section 4: 
%
%%%%%%%%%%%%%%%%%%%%%%%%%%%%%%%%%%%%%%%%%%%%%%%%%%%%%%%%%%%%%%%%%%%%%%%%%%%%%%%%%%%%%%%%%%%%%%%%%%%%%%%%%%%%%%%
%
\section{Spontaneous Symmetry Breaking, Dissipation and Geometric phase}
\label{sec:SSB,Dissipation,GP}
For a finite magnetic field $h\neq 0$, the pole contribution to magnetization is given in the momentum representation by
\begin{eqnarray}
\label{pseudoNG}
M(h)=-2\mu_{B}h\left(\frac{\chi(q^2,h)}{q_{0}-\omega(\mathbf{q},h)+i\varepsilon}\right)_{q_{0}=0,{\mathbf q=0}},
\end{eqnarray}
with $\lim_{{\bf{q}}\to 0}\omega(\mathbf{q},h)=2\mu_{B}h$, and $\chi(0,h)=M(h)$. This means that in an external field the magnon becomes massive. This is a common property of pseudo NG bosons\cite{UMWA1}.

Let us assume now that the external field $h$ is changing with time. Integrating both sides of equation (\ref{pseudoNG}) one gets
\begin{eqnarray}
\label{contint}
\int M(h)d h=\int\chi(0,h)d h .
\end{eqnarray}
This essentially agrees with the result of Jain and Pati \cite{JainPati}, if one recalls that in the linear response theory the imaginary part of the response function (the susceptibility) corresponds to absorption \cite{Lovesey}. In our case all the response at $\mathbf{q}=0$ is due to a simple pole in the response function  
\begin{eqnarray}
\chi(0,h)=-Im \int_{-\infty}^{+\infty}\frac{ d q_{0}}{2\pi}\left(\frac{\chi(q^2,h)}{q_{0}-\omega(\mathbf{q},h)+i\varepsilon}\right)_{\mathbf q=0}
\end{eqnarray}

However there is also an important difference. In our appoach it suffices to couple a one-dimensional external field to the system, the second degree of freedom being provided by the order parameter itself. Thus no rotation of the external field is needed to generate the geometric phase. This result is satisfactory, since hysteresis in ferromagnets is observed for a magnetic source whose magnitude is varying  but whose direction is fixed. 

Now we show that, when the system is in a broken symmetry ground state and the integral (\ref{contint}) is evaluated for a certain closed contour in $h-M$ plane, that integral is non-vanishing. In order to find the functional relation of magnetization to the external field $M(h)$, we need to evaluate the Green's functions of quasiparticles in the system. Here we calculate the Green's functions approximately by truncating the (infinite) chain of WT-identities and solving the resulting system of equations. The lowest order approximation is achieved by truncating the WT-chain of identities at the level of the vertex-propagator identity
\begin{eqnarray}
S_{\uparrow}^{-1}(p)-S_{\downarrow}^{-1}(p)&=&2\mu_{B}h-M\Gamma(p,-p,0).
\end{eqnarray}
If one approximates the vertex with a constant $\lambda$, one can determine the approximate boson and fermion propagators consistent with the WT-identities. This corresponds to the Hartree-Fock (HF) approximation for the fermions and the random-phase-approximation (RPA) for the boson (magnon) field. The approximate fermionic propagators are
\begin{eqnarray}
S^{HF}_{\uparrow}(p)&=&\frac{1}{p_{0}-(\epsilon_p-\mu)+\frac{1}{2}(2\mu_{B}h -\lambda M)+i\varepsilon}\\
S^{HF}_{\downarrow}(p)&=&\frac{1}{p_{0}-(\epsilon_{p}-\mu)-\frac{1}{2}(2\mu_{B}h -\lambda M)+i\varepsilon} .
\end{eqnarray}
where $\mu $ is the chemical potenitial of the interacting system, and $\epsilon_p= p^2/(2m)$.
Note that this structure of the spin-up and down electron propagators follows from the spontaneous $SU(2) \rightarrow U(1)$ symmetry breaking. This structure is observed in experiments as a characteristic splitting in the dispersion relations of the spin-up and down electrons \cite{White}.

Having found the Green's functions of the single-particle states in HF-approximatation, we now can estimate the densities of spin-up and down electrons. For the majority spin $n_{\uparrow}$ we get 
\begin{eqnarray}
\label{updensity}
n_{\uparrow}&=&-i\lim_{\eta\to 0}\int \int \frac{dp_o}{2\pi}\frac{d^3 p}{(2\pi)^3}S^{HF}_{\uparrow}(p_o,p)e^{-i\eta p_o} \nonumber \\
            &=&\int \Theta(-\epsilon_p+\mu-\frac{1}{2}(2\mu_B h-\lambda M))\frac{d^3 p} {(2\pi)^3}        \nonumber \\
            &=&\frac{n}{{2\varepsilon_F}^{3/2}}(\mu- \frac{1}{2}(2\mu_B h -\lambda M))^{3/2}.
\end{eqnarray}
and similarly for the minority spin $n_{\downarrow}$:
\begin{eqnarray}
\label{downdensity}
n_{\downarrow}=\frac{n}{{2\varepsilon_F}^{3/2}}(\mu+\frac{1}{2}(2\mu_B h -\lambda M))^{3/2}.
\end{eqnarray}
One needs to determine $\mu$, and $M$, subject to the following constraints:
\begin{eqnarray}
\label{constraints}
n=n_{\uparrow}+n_{\downarrow} \nonumber\\
M=n_{\uparrow}-n_{\downarrow}.                              
\end{eqnarray}
Eqs. (\ref{updensity})-(\ref{constraints}) lead to
\begin{widetext}
\begin{eqnarray}
n&=&\frac{n}{{2\varepsilon_F}^{3/2}}\left((\mu- \frac{1}{2}(2\mu_B h -\lambda M))^{3/2}+
                                            (\mu+\frac{1}{2}(2\mu_B h -\lambda M))^{3/2}\right)\\
M&=&\frac{n}{{2\varepsilon_F}^{3/2}}\left((\mu- \frac{1}{2}(2\mu_B h -\lambda M))^{3/2}-
                                            (\mu+\frac{1}{2}(2\mu_B h -\lambda M))^{3/2}\right).                             
\end{eqnarray}
\end{widetext}
Now using the definitions in eq.(\ref{definitions}) these conditions can be rewritten as 
\begin{eqnarray}
1+\Delta&=&\frac{1}{{\varepsilon_F}^{3/2}}(\mu- \frac{1}{2}(2\mu_B h -\lambda M))^{3/2}\\
1-\Delta&=&\frac{1}{{\varepsilon_F}^{3/2}}(\mu+ \frac{1}{2}(2\mu_B h -\lambda M))^{3/2}.                         
\end{eqnarray}
We thus arrive at an equation linking the magnetization to the applied field
\begin{eqnarray}
\label{orderh}
(1-\Delta)^{2/3}-(1+\Delta)^{2/3}=\frac{2\mu_Bh-\lambda n\Delta}{\varepsilon_F}.                     
\end{eqnarray}
Note that for $h=0$ it reduces to the previously obtained variational condition that determines spontaneous magnetization. Thus eq.(\ref{orderh}) represents a generalization of eq.(\ref{Stoner}) for a case of non zero external field $h$. One can also obtain this condition directly by fixing $h$ and minimizing the ground state energy in HF-approximation with respect to $\Delta$
\begin{eqnarray}
E_{HF}(\Delta,h)=\frac{3}{10}\varepsilon_{F}n((1+\Delta)^{5/3}+(1-\Delta)^{5/3})\nonumber \\
                                            +\frac{\lambda n^2}{4}(1-\Delta ^2)+\mu_Bnh\Delta .
\end{eqnarray}
%\vspace{4in}
%%%%%%%%%%%%%%%%%%%%%%%%%%%%%%%%%%%%%%%%%%%%%%%%%%%%%%%%%%%%
\begin{figure}[hbt]
\includegraphics[width=2.5in]{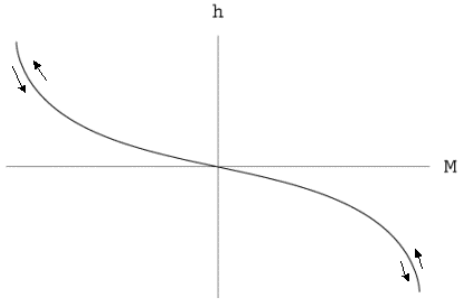}
\caption{Driven magnetization $M(h)$ in a symmetric ground state}
\label{valley_bed1}
\end{figure}
%%%%%%%%%%%%%%%%%%%%%%%%%%%%%%%%%%%%%%%%%%%%%%%%%%%%%%%%%%%%
%%%%%%%%%%%%%%%%%%%%%%%%%%%%%%%%%%%%%%%%%%%%%%%%%%%%%%%%%%%%
\begin{figure}[hbt]
\includegraphics[width=2.5in]{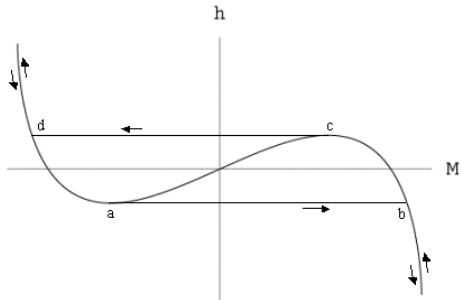}
\caption{Driven magnetization $M(h)$ in an asymmetric ground state }
\label{valley_bed2}
\end{figure}
%%%%%%%%%%%%%%%%%%%%%%%%%%%%%%%%%%%%%%%%%%%%%%%%%%%%%%%%%%%%
%
%
The dependence of magnetization on $h$ is illustrated in FIG.3 and FIG.4 for the two cases: a) the system does not display spontaneous magnetization in its ground state  and b) the system gets spontaneously magnetized in the absence of external field i.e. when $ 2/3<x< 2^{-1/3}$.
%
%
%%%%%%%%%%%%%%%%%%%%%%%%%%%%%%%%%%%%%%%%%%%%%%%%%%%%%%%%%%%%
%
\begin{figure}[hbt]
\includegraphics[width=2.5in]{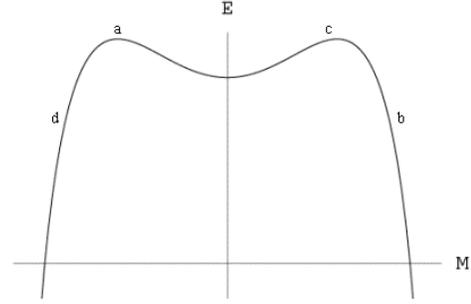}
\caption{Ground state energy along the valley of stability as a function of magnetizarion}
\label{valley_bed}
\end{figure}
%%%%%%%%%%%%%%%%%%%%%%%%%%%%%%%%%%%%%%%%%%%%%%%%%%%%%%%%%%%%

One can relate the characteristic s-shape of the curve $M(h)$ to the form of the meandering symmetric valley of the minimum ground state energy. Although the bed of this valley remains a local minimum for fixed $h$, this bed does not stay at a constant level but rises and falls as the applied field changes. One can express $h$ through $M$ from eq.(\ref{orderh}), and substitute that expression in $E(h(M),M)$. The resulting level of the valley as a function of $M$ is given in FIG.5. We see that at the points on the hysteresis curve where $M$ jumps ($a\rightarrow b$ and $c\rightarrow d$) the destination points have lower energy than the starting points. This indicates that directional tunneling (in reality also thermally assisted) between these states may start ahead of the turning points, as often observed in the experimental hysteretic curves.

One can see from these pictures that a system that displays an instability of the ground state with respect to spontaneous magnetization, behaves very differently from a system that does not have such instability. The former displays a hysterisis loop when the external field is varied in a cycle, while the latter does not. The area of the hysterisis loop gives the dissipated magnetic energy. In other words we have proven that 
\begin{eqnarray}
\oint M(h)d h=\oint\chi(0,h)d h\neq 0 .
\end{eqnarray}
This area gives the cyclic Jain-Pati dissipative geometric phase in our model.
The result can easily be extended to finite temparature. The dominant part of the low-energy spin response at finite temperature again comes from the magnon pole at $2\mu_{B} h$, but now its spectral function is temperature dependent \cite{UMWA1},\cite{MUST},\cite{Whitehead} 
\begin{eqnarray}
M(h,T)=Im \int_{-\infty}^{+\infty}\frac{ d q_{0}}{2\pi}\left(\frac{\chi(q^2,h,T)}{q_{0}-\omega(\mathbf{q},h)+i\varepsilon}\right)_{\mathbf q=0} 
\end{eqnarray}
This suggests that $M(h,T)$ can be determined from a finite temperature extention of eq.(\ref{orderh}) using the standard technique described in \cite{Animalu}. We first define 
\begin{eqnarray}
\varepsilon^{\uparrow}_p\equiv\varepsilon_p+\frac{1}{2}(2\mu_B h-\lambda M)\\
\varepsilon^{\downarrow}_p\equiv\varepsilon_p-\frac{1}{2}(2\mu_B h-\lambda M).
\end{eqnarray}
Then the spin-up electron density is 
\begin{eqnarray}
n_{\uparrow}&=&-i\lim_{\eta\to 0}\int \int\frac{d^4 p}{(2\pi)^4}
                 \left(\frac{S^{HF}_{\uparrow}(p_o,p)}{1+\exp\beta(\mu-\varepsilon^{\uparrow}_p)}
                                                                                     \right)e^{-i\eta p_o}\nonumber\\
&=&\frac{(2m)^{2/3}}{4\pi^2}\int_{0}^{\infty}\frac{\varepsilon^{1/2}}{1+\exp(\beta(\mu-\varepsilon-\frac{1}{2}(2\mu_B h-\lambda M))}d\varepsilon\nonumber\\
&=&\frac{(2m)^{2/3}}{4\pi^2}(kT)^{3/2}\int_{0}^{\infty}\frac{x^{1/2}}{1+\exp(\alpha(M,h)-x)}d x ,
\end{eqnarray}
with 
\begin{eqnarray}
\beta=1/kT\quad\quad
x\equiv\beta\varepsilon\quad\quad\quad
\alpha(M,h)\equiv\beta(\mu-\mu_{B}h+\lambda M/2) .\nonumber
\end{eqnarray}
Similarly the spin-down electron density is 
\begin{eqnarray}
n_{\downarrow}=\frac{(2m)^{2/3}}{4\pi^2}(kT)^{3/2}\int_{0}^{\infty}\frac{x^{1/2}}{1+\exp(\gamma(M,h)-x)}d x ,
\end{eqnarray}
with 
\begin{eqnarray}
\gamma(M,h)\equiv\beta(\mu+\mu_{B}h-\lambda M/2).
\end{eqnarray}
Defining
\begin{eqnarray}
F(\alpha)= \int_{0}^{\infty}\frac{x^{1/2}}{1+\exp(\alpha(M,h)-x)}d x ,
\end{eqnarray}
one finds 
\begin{eqnarray}
M= \frac{(2m)^{2/3}}{4\pi^2}(kT)^{3/2}(F(\alpha(M,h))-F(\gamma(M,h)))\nonumber\\
n=\frac{(2m)^{2/3}}{4\pi^2}(kT)^{3/2}(F(\alpha(M,h))+F(\gamma(M,h))) ,
\end{eqnarray}
from which it follows 
\begin{eqnarray}
M= n\left(\frac{(F(\alpha(M,h))-F(\gamma(M,h))}{(F(\alpha(M,h))+F(\gamma(M,h))}\right) .
\end{eqnarray}
One can in principle determine both $M(h,T)$ and $\mu(h,T)$ from these equations by solving them numerically. Then to compute the finite-temperature Berry's phase one simply needs to integrate the finite-temperature order parameter along the hysteresis loop. There is nothing surprising about this result. We have simply come to a reinterpretation of the first law of thermodynamics as a balance of dynamic and geometric phases in a many-body system
\begin{eqnarray}
TdS=dU+pdV-\vec{H}\cdot\vec{dB}
\end{eqnarray}
As often seen before, the geometric phase gives us a new perspective on a classical result.

\section{Conclusions}
We have shown that when a quantum many-body system with a spontaneous symmetry breaking in the ground state evolves in time in such a way that an order parameter completes a cycle, one can associate a certain geometric phase with the energy dissipated in this cycle. We have examined the Stoner model of itinerant ferromagnetism and have shown that when the system has ferromagnetic instability, i.e., spontaneous magnetization occurs in the absence of applied external field, then hysteresis arises with respect to a changing external field and the magnetization process is irreversible. Thus, spontaneous symmetry breaking seems to have bearing upon another important concept in physics - reversibility. The measure of irreversibility is given by the dissipation, i.e., the dissipative geometric phase. 
Our preliminary investigation has confirmed that a similar mechanism is responsible for the hysteresis observed in other many-body systems displaying spontaneously broken symmetry in the ground state. We plan to address this issue in a forthcoming publication.

\section{Acknowledgements}
I am indebted to the late Prof. H. Umezawa for introducing me to the techniques that he and his collaborators developed and employed to many-body problems over the years. I wish to thank Prof. Pawel O. Mazur for encouraging me to publish this paper. I wish to acknowledge stimulating discussions with the late Prof. Jeeva Anandan, Dr. Alonso Botero, Jun Suzuki and Piruz Vargha and to thank them  and for their advice. I also wish to thank Prof. Kuniharu Kubodera for a careful reading of the manuscript. This work is supported in part by the National Science Foundation, Grant No. PHY-0140214.

\end{document}